\documentclass[prl, twocolumn]{revtex4}
\usepackage{graphicx}
\usepackage{hyperref}
\usepackage{amsmath}

\begin{document}

\title{Trans-spectral orbital angular momentum transfer via four wave mixing in Rb vapor}

\author{G.\ Walker,$^{1}$ A.\ S.\ Arnold,$^{2}$ and S.\ Franke-Arnold$^{1}$} \affiliation{$^{1}$School of Physics and Astronomy, SUPA, University of Glasgow, Glasgow G12 8QQ, UK}
\affiliation{$^{2}$Dept.\ of Physics, SUPA,  University of Strathclyde, Glasgow G4 0NG, UK}

\date{\today}

\begin{abstract}
We report the transfer of phase structure, and in particular of orbital angular momentum, from near-infrared pump light to blue light generated in a four-wave-mixing process in $^{85}$Rb vapour.  The intensity and phase profile of the two pump lasers at $780\,$nm and $776\,$nm, shaped by a spatial light modulator, influences the phase and intensity profile of  light at $420\,$nm which is generated in a subsequent coherent cascade.  In particular we oberve that the phase profile associated with orbital angular momentum is transferred entirely from the pump light to the blue.  Pumping with more complicated light profiles results in the excitation of spatial modes in the blue that depend strongly on phase-matching, thus demonstrating the parametric nature of the mode transfer.  These results have implications on the inscription and storage of phase-information in atomic gases.
\end{abstract}

\maketitle


Phase-matched nonlinear optical processes include a wide range of phenomena including spontaneous parametric down-conversion (SPDC), four-wave mixing (FWM), and stimulated Raman scattering. They can be studied in dielectrics as well as in atomic media. Phase-matching, if applied to the longitudinal component of the wavevector, results in momentum correlations between the relevant light fields but it also allows the transfer of phase information. 
The spatially varying phase of Laguerre-Gauss (LG) modes is particularly appealing when investigating phase-dependent effects, as these eigenmodes of propagation are characterised by an axial vortex line which can easily be visualised.  LG modes have an azimuthal phase structure of $\exp(i l \varphi)$ associated with an orbital angular momentum (OAM) of $l \hbar$ per photon \cite{Allen92}. 
Conservation and storage of OAM has been observed in a number of processes including the entanglement of OAM modes in SPDC \cite{Mair01, Caetano02}, second harmonic generation \cite{Courtial97}, or FWM in semiconductors \cite{Ueno09}.  For SPDC it has been shown that phase-matching applied to waves with a transverse phase-structure is responsible for the conservation of OAM \cite{Franke-Arnold01} as long as all modes are observed \cite{torres}.  More generally, parametric amplification of complex images has been realised using non-linear crystals \cite{image}.

In atom optics, unlike in solid state processes, nonlinear effects are highly efficient and require only low light intensities.
FWM and other closed loop processes have enabled the transfer of phase information to room temperature atomic vapours \cite{Davidson,Boyer08}, and optical OAM has been used to manipulate the quantum state of Bose-Einstein condensates via Raman transitions \cite{Andersen06, Brachmann11}.
In these processes, the OAM of light was transferred to the atomic media.  In a recent experiment, correlated OAM beams, were generated in a FWM process utilising Raman transitions between different hyperfine states of the same optical transition \cite{lett}.  
Here we report a FWM process that transfers the OAM of the pump light  -- via the atomic medium -- to light at the opposite end of the visible spectrum.  

\begin{figure}[!b]
\centering\includegraphics[width=\columnwidth]{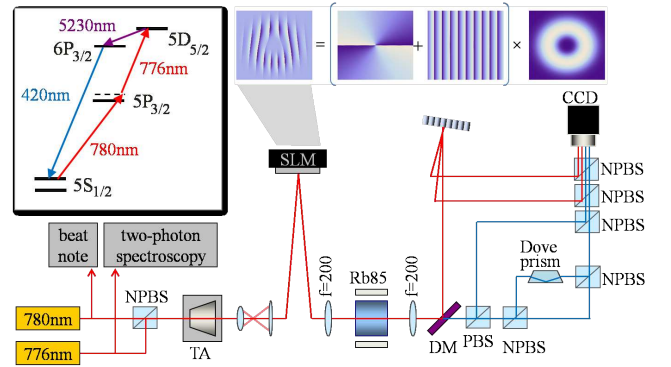}
\caption{\label{setup} Experimental set-up, with abbreviations: N/PBS (non-/polarizing beamsplitter), DM (dichroic mirror), TA (tapered amplifier), SLM (spatial light modulator) and CCD (charge-coupled device).  Inset:  $^{85}$Rb FWM level scheme.}
\end{figure}

We recently demonstrated the highly efficient generation of up to 1.1mW of coherent $420\,$nm blue light by enhanced frequency up-conversion in Rb vapour \cite{Vernier10}, while similar experiments have been performed by a number of other groups in Rb \cite{Zibrov02, Meijer06, Akulshin09} and Cs \cite{Schultz09}. The blue light generation relies on two near-infrared (NIR) pump lasers at $780\,$nm ($5S_{1/2}$--$5P_{3/2}$) and $776\,$nm ($5P_{3/2}$--$5D_{5/2}$).  The atoms can subsequently decay in a cascade via $6P_{3/2},$ generating coherent infrared ($5.23\,\mu$m) and blue ($420\,$nm) light ($^{85}$Rb level scheme in Fig \ref{setup}).  
We investigate phase-matching in the context of the FWM process by illuminating the atomic medium with pump light in various LG modes and some of their simple superpositions. 

Any image, including the profile of our pump and generated light beams, can be described in terms of a superposition of LG modes $\Phi_p^l(x,y,z)$, where $p$ and $l$ denote the number of radial nodes and the winding number, respectively. The normalised LG modes are given by:
\begin{eqnarray}
\Phi_p^l &=& \sqrt{\frac{2p!}{\pi \left(|l|+p\right)!}}\,\frac{1}{w(z)}\left(\!\frac{\sqrt{2}r}{w(z)}\!\right)^{\!\!|l|}\, L_{p}^{|l|}\!\!\left[\frac{2r^{2}}{w^{2}(z)}\right] {\rm e}^{il\varphi}
\;\nonumber \\
& & \times {\rm e}^{-r^2/w^2} \; {\rm e}^{+i(2p+|l|+1)\arctan(z/z_R)} {\rm e}^{-i k z}. \label{e1}
\end{eqnarray}
Here $L_{p}^{| l |}$ is an associated Laguerre polynomial and $w(z)=w_0 \sqrt{1+z^2/z_R^2}$ for a beam waist $w_0$, with a Rayleigh range $z_R=\pi {w_0}^2/\lambda$.
For simplicity we restrict ourselves to images represented by pure LG modes and their simple superpositions in order to show that intensity and phase information can be transferred via a FWM process from the pump light to the generated blue light. 

All theoretical considerations are based on a simple phase-matching approach, without including a full description of the propagation equations, and in particular effects due to absorption and Kerr lensing of driving and generated fields which become important in the vicinity of single photon resonances \cite{Vernier10}.  

FWM leads to the generation of blue and infrared light under four photon resonance.  The efficiency of the process is determined by the overlap of the four light amplitudes of the two pump lasers and the generated blue (B) and infrared (IR) light, $E^*_{780} E^*_{776} E_{\rm B} E_{\rm IR}$. In our experiment, the mode profiles of the input lasers are chosen by diffraction off a spatial light modulator (SLM), and we only investigate pump modes with $p=0.$  Any output mode profile can be represented as a suitable superposition of LG modes with a given waist $w_0$.  

The probability amplitude to generate a specific combination of a blue and an infrared photon in modes $\Phi_{p_{\rm B}}^{l_{\rm B}}$ and  $\Phi_{p_{\rm IR}}^{l_{\rm IR}}$ is given by the overlap with the pump profiles,
\begin{equation}
C_{p_{\rm B}, p_{\rm IR}}^{l_{\rm B}, l_{\rm IR}} = \int_{-L/2}^{L/2} \int_{0}^{R} \!\!\!\! \int_{0}^{2 \pi} \!\!\!\! {\rho \,  {\Phi_{p_{780}}^{l_{780}}}^\ast  {\Phi_{p_{776}}^{l_{776}}}^\ast  \Phi_{p_{\rm B}}^{l_{\rm B}} \Phi_{p_{\rm IR}}^{l_{\rm IR}}  d\varphi \, d\rho \, dz},  \label{e2}
\end{equation} 
where $L$  is the length of the vapour  cell and $R$ the numerical aperture of the system.  The probability that  blue and IR photons are in the respective modes is then  $|C_{p_{\rm B}, p_{\rm IR}}^{l_{\rm B},l_{\rm IR}}|^2$.  We note an interesting similarity with pattern formation, where certain superpositions of OAM modes are favoured due to their overlap integral \cite{grynberg}.

Inserting the LG modes (\ref{e1}), and integrating over the azimuthal angle leads to 
\begin{equation}
\int_{0}^{2 \pi} {{\rm e}^{i (l_{776}+l_{780}-l_{\rm B}-l_{\rm IR})\varphi} d\varphi}=2 \pi \delta_{l_{776}+l_{780}-l_{\rm B}-l_{\rm IR}},\nonumber
\end{equation}
where $\delta$ is the Kronecker delta. This restricts the output modes to those for which the OAM is conserved,  $$ l_{776}+l_{780}=l_{\rm B}+l_{\rm IR}. $$  

A further condition for the output modes results because the largest conversion efficiency requires matched Gouy phases, which is mathematically linked to the integration over $z$ in (\ref{e2}).  For Gaussian beams the well-known Boyd criterium \cite{Boyd68} states that the conversion efficiency in frequency translation processes is maximised when the Rayleigh ranges $z_{R}$ of all involved beams  are equal.  In usual SPDC set-ups, the Rayleigh range far exceeds the crystal length, removing the need for Gouy phase matching, whereas the length of our vapour cell is many Rayleigh ranges ($25\,$mm). Applied to our FWM scheme, the Boyd criterium implies $w_{\rm B} = w_{780}\sqrt{420/780},$ $w_{\rm IR} = w_{780}\sqrt{5230/780},$ making the IR beam waist about 3.5 times larger than that of the blue beam. However, in general the Gouy phase depends on the mode number, and for LG modes it is given by $(2p+|l|+1) \arctan(z/z_R)$.  For matched Rayleigh ranges, and assuming zero order radial modes for the pump light, this results in maximum efficiency if $$|l_{780}|+|l_{776}|=|l_{\rm B}|+|l_{\rm IR}|+2p_{\rm B}+2p_{\rm IR}.$$ 

The IR beam is unobserved in the experiment, but OAM conservation and Gouy phase matching indicate it is mainly generated in the fundamental Gaussian mode.  The components of the blue light in a given  mode $\Phi_{p_{\rm B}}^{l_{\rm B}}$  are found by evaluating the overlap integral (\ref{e2}) for all allowed mode combinations.  This becomes important when investigating pump light in superposition modes.

A simplified setup of the experiment is shown in Fig.\ref{setup}. Pump light at $780\,$nm and $776\,$nm was generated by external cavity diode lasers \cite{arnold}, overlapped at a non-polarising beamsplitter and fibre-coupled to a 
tapered amplifier (TA). The maximum output power of approximately $500\,$mW allowed strong nonlinear coupling, and the fibre coupled TA output ensured clean Gaussian pump modes.  

The frequency of the $780\,$nm light was determined from its beatnote with another $780\,$nm laser locked to the $F = 3$ to $F' =3, 4$  cross-over transition.  The $776\,$nm frequency was measured with counter-propagating two photon spectroscopy in a further heated Rb cell, 
where the absorption of the $776\,$nm laser could be determined relative to the saturated absorption trace of the $780\,$nm laser.  The frequency could then be set manually by adjusting the voltage across a piezo electric device controlling the external grating in Littrow configuration.   

The overlapped pump lasers were converted to LG modes or superpositions thereof by diffraction off a suitable hologram displayed on an SLM (Hamamatsu X10468).  The computer generated holograms  consist of the required phase of the desired mode superimposed with that of a linear diffraction grating, generating a fork dislocation as indicated in Fig.\ \ref{setup} for a $\Phi_0^2$ mode.  Although the SLM operates on phase only, the intensity can be shaped by modulating the amplitude of the grating.  
The spatially sculpted pump light was focussed into a Rb cell heated to $120^{\circ}$C (a vapour pressure of $9\times 10^{-4}$mbar) and after the cell was split from the generated blue light by a dichroic mirror. The generated IR light was absorbed by the cell windows.  The remaining NIR pump beams were separated using a diffraction grating followed by a propagation distance of  $\approx 2.5\,$m.

The dense rubidium vapour acts as a strong nonlinear material and the beams are subject to Kerr lensing -- the change of refractive index with input intensity. This is particularly apparent in the $780\,$nm beam 
shows strong self-focussing and defocussing depending on the detuning. As this can negatively affect the amplitude overlap of the four light fields, especially when using spatially shaped pump lasers, all data has been taken at two-photon resonance and at minimum Kerr lensing. In a preliminary experiment we have verified that detunings of $\Delta_{780} = -\Delta_{776}$ = 1.6GHz \cite{Vernier10} indeed yield minimal Kerr lensing for Gaussian pump modes, but shift to slightly higher frequencies for higher order LG modes.  Using a simple diffraction grating, rather than a fork hologram, we monitored the transmission of the Gaussian pump lasers through the Rb vapour cell, and adjusted the pump frequency combination to yield minimal Kerr lensing which is particularly noticable for the  $780\,$nm transition.  Once set the pump frequencies stayed stable throughout the experiment.  

\begin{figure}[!b]
\centering\includegraphics[width=.9 \columnwidth]{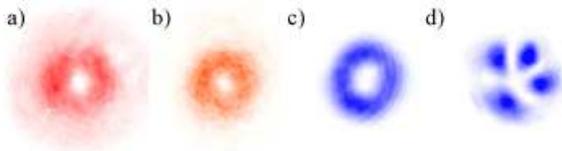}
\caption{\label{LGdata}  Evidence of OAM transfer: Example CCD data at maximum blue intensity and minimum Kerr effects for $\Phi_0^1$ input modes at  $780\,$nm (a) and  $776\,$nm (b). Resulting intensity profile (c) and interferogram (d) at $420\,$nm. Both c) and d) are consistent with a blue $\Phi_0^2$ mode. }
\end{figure}

For the first part of the work, both  input laser beams were shaped as LG beams with $p=0$ and the same azimuthal number  $l_{\rm NIR}=l_{780}=l_{776}$  which was varied between $0$ and $5$.  An example of both pump output profiles shaped as $\Phi_0^1$  is displayed in Figs.\ \ref{LGdata}a and b.
In order to visualise both the intensity and phase profile of the generated blue light, a third of the light was monitored on a CCD camera (Fig.\ \ref{LGdata}c).  The remaining blue light passed through a Mach-Zehnder interferometer containing a Dove prism in one of the arms. The Dove prism flips the image in this interferometer arm, thereby effectively reversing the OAM of the light. For blue light with a particular OAM $l_{\rm B},$ the resulting interference pattern then contains a superposition of modes with $\pm l_{\rm B}$ \cite{FrankeArnold07}, exhibiting the characteristic 2$l_{\rm B}$ lobes 
shown in Fig.\ \ref{LGdata}d.  

\begin{figure}[!b]
\centering\includegraphics[width=.9 \columnwidth]{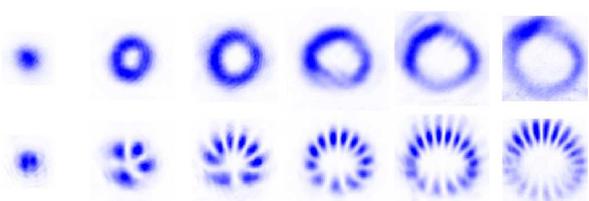}
\caption{\label{LGresults} OAM transfer from IR pump light with $l_{\rm NIR}=0 \to 5$ to blue light at $420\,$nm beams, with intensity profile (top) and  interferograms (bottom) consistent with $l_{\rm B}=l_{780}+l_{776}$.}
\end{figure}

The blue intensity and interference profile resulting from input modes  $l_{\rm NIR}$ = 0 to 5 is shown in Fig.\ \ref{LGresults}.  Comparing the intensity profile with the theoretical profile of pure LG modes, we find that each blue mode has an OAM number of $l_{\rm B}= 2 l_{\rm NIR}.$   This is confirmed by the interference fringes, which display 0 to 20 lobes. This indicates that the IR light is generated in the mode $\Phi_0^0,$ which has the largest overlap with the NIR input modes.  All of the OAM is transferred into the blue beam with no observable contribution in other OAM states. 

In the second part of our work we  investigated the phase coherence of the parametric process, showing that OAM conservation is inherently quantum in nature.   For the pure LG modes one may argue that the blue output modes are solely determined by the intensity overlap with the input light.  Instead we find that interference between the excitation amplitudes determines the output modes.  This becomes apparent by shaping the pump beam profiles as superpositions of two LG modes, with OAM values $l$ and $m$ \cite{FrankeArnold07}.
The combined input field is
$$ E_{780} E_{776}=\frac{1}{2}(\Phi_0^{l} + \Phi_0^{m})_{780}  (\Phi_0^{l} + \Phi_0^{m})_{776},$$ 
with multiple absorption paths in the initial two-photon absorption, allowing a potential OAM transfer of $2l,$ $2m,$ or $m+l.$  Again, assuming that no OAM is transmitted to the infrared beam, $ E_{\rm IR}= (\Phi_0^0)_{\rm IR}$, and for modes with matching Gouy phases, we expect a blue output mode
\begin{equation} \label{modes}
 E_{\rm B}=\left(u_{2l}\Phi_0^{2l} +u_{2m} \Phi_0^{2m}+u_{l+m}\Phi^{m+l}_{|m|+|l|-|m+l|/2}\right)_{\rm B} , 
\end{equation}
where the complex amplitudes $u_i$ depend on the mode-overlap as calculated from (\ref{e2}). 

For this part of the experiment the TA was removed from the set-up to further limit the amount of Kerr lensing.  We also omitted the Mach-Zehnder interferometer, because phase structure is inherently present in the overlap of different superposition modes. 

\begin{figure}[!h]
\centering\includegraphics[width=\columnwidth]{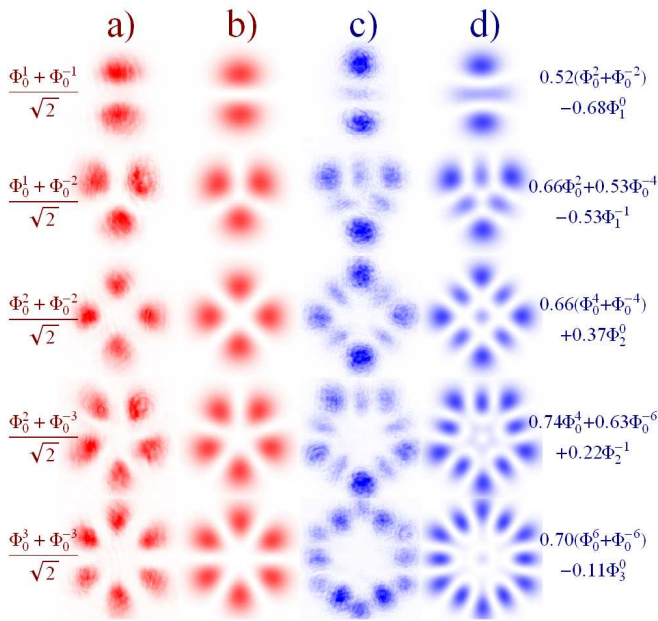}
\caption{\label{Superposition} Phase transfer of superposition modes:  Experimental realisation of the pump superposition modes detailed as red text (a) and corresponding observed blue output modes (c).  Theoretical simulation of pump modes (b) and anticipated output modes detailed as blue text (d). }
\end{figure}

Fig.\ \ref{Superposition} shows the observed profile of the pump modes set to 5 different superpositions (in column a), and the generated blue light (in column c).  Simulations of the pump modes and the anticipated blue modes are shown in columns b) and d).  Here, the anticipated composition of the blue light was evaluated by demanding OAM conservation and Gouy phase matching (\ref{modes}), and is shown in blue text.

We find good agreement between experiment and simulation, confirming the importance of OAM conservation and matched Gouy phases for the studied FWM process.  In particular we find that the generated mode patterns cannot be explained simply from the intensity overlap of pump and generated light, but that phase-matching and amplitude interference play a crucial role.  We always observe a non-zero $p$ component in the mode corresponding to a transfer of $l+m$ units of OAM from pump to blue light, corresponding to an excitation of $l\hbar$ on the $780\,$nm transition and $m\hbar$ on the $776\,$nm transition, or vice versa.  Just considering the excitation probabilities, the interference mode $m+l$ should be excited with a probability of $1/2$, whereas the modes with $2l$ and $2m$ should occur with a probability of $1/4$ each.  The actual excitation probability deviates from this, as the FWM process depends on the mode-overlap in the vapour.


In conclusion we have observed the transfer of the combined OAM of the NIR pump light to the coherent blue light at $420\,$nm via FWM in a $^{85}$Rb vapour, leaving the unobserved IR transition in the fundamental Gaussian mode.  We have demonstrated this for pump beams in LG modes with $l_{\rm NIR} = 0 \to 5$, which generated blue LG modes with $l_{\rm B}=2 l_{\rm NIR}$.  We explain the fact that the OAM is not split between the  generated blue and IR light by the large difference in waists, which maximises the mode overlap in the FWM process for all OAM in the blue.     

Through the use of simple superpositions of LG beams as input modes we have illustrated the parametric nature of the process.  We have noted that Gouy phase matching of LG modes determines not only the ideal beam waists of the generated modes, but also favours modes that conserve mode number.  We have successfully predicted the output modes as those which obey OAM conservation and have matched Gouy phases. A more detailed description would require the propagation of the light modes, in particular in cases where Kerr lensing is not negligible, and experimental access to the generated IR light.

The observed effects are not restricted to light carrying OAM but should apply to all phase information inherent in the pump light. The described process should allow the conversion of phase information from a given input frequency to a completely different frequency.  

We are grateful for useful discussions with Erling Riis and preliminary experiments by Aline Vernier.  SFA acknowledges financial support by the European Commission via the FET Open grant agreement Phorbitech FP7-ICT-255914.

\end{document}